# Stochastic Parrots or Singing in Harmony? Testing Leading AI LLMs' Ability to Replicate a Human Survey with Synthetic Data

Jason Miklian, Research Professor (corresponding)

University of Oslo; jason.miklian@globe.uio.no

Kristian Hoelscher, Research Director

Peace Research Institute Oslo; krihoe@prio.org

John Katsos, Professor

American University of Sharjah; jkatsos@aus.edu

**Abstract**

How well can AI-generated synthetic research data replicate the responses of human participants? This article benchmarks synthetic survey generation against a survey of 400 Silicon Valley coders and developers by asking first for leading consumer AI platforms to produce full synthetic datasets through a naturalistic chat protocol, then testing the three newest frontier models (Claude Fable 5, GPT-5.6 Sol, and Kimi K3). Our findings reveal that while the platforms produced technically plausible results that lean more towards replicability and harmonization than typically assumed, none of the LLMs consistently captured the complete set of counterintuitive findings of the human survey. This pattern varied across different prompt formations and generation methods: deviations in synthetic data grouped together for some model comparisons, while others diverged substantially. While leading LLMs are increasingly being used to scale, replicate and replace human survey responses in innovation and science policy research, the tested models still differed substantially from the human benchmark. Agreement among synthetic samples, whether across models or across runs, therefore does not establish agreement with human populations, showing here that replicability and validity decouple. We propose that synthetic respondents could be used before fieldwork, as an instrument for identifying model-generated expectations about research populations, and we connect this use to emerging calibration methods for settings where a human anchor exists.

## 1. Introduction

Survey research can contribute invaluable richness and depth to empirical and theoretical scholarship. But it faces two persistent structural challenges: time and cost (Davidson and Karell, 2025). This burden can pressure scholars as funding streams diminish and the "publish-or-perish" criteria for tenure and promotion become more intense. Yet demand for depth is outrunning what current research designs can deliver, especially in understanding the social impacts of organizations (Hayes, 2025). If months of survey fieldwork can be accomplished in a fraction of the time and cost, it may present an opportunity for innovation and science policy researchers and those who rely on their findings. The wide-scale proliferation of powerful Generative Artificial Intelligence (GenAI) tools that use probabilistic large language models (LLMs) to generate human-like survey responses raises four inter-connected questions: Could synthetic survey responses generated by AI systems create data that is as plausible as real human responses in complex social contexts? If so, is this synthetic material analytically useful? Does it merely reflect conventional wisdom without uncovering new truths? And does the answer depend on how the synthetic sample is produced?

The literature on silicon sampling has largely treated generation method as an implementation detail, with a standard emerging around respondent-level generation in which each synthetic participant is produced independently, conditioned on a demographic profile (Argyle et al. 2023; Ashokkumar et al. 2026). Researchers in practice, though, increasingly generate synthetic data the way anyone with a subscription would: by pasting a survey into a chat window and asking for completes. Whether these routes yield the same answers is an empirical question that, to our knowledge, no study has tested against a human benchmark whose availability in model training data is unknown.

The intersecting pressures for more detailed research data and the questions that GenAI tools raise creates both opportunities and risks. One opportunity lies in AI's potential to augment research capabilities to assist with coding, identify patterns across large datasets, and help manage increasingly complex information landscapes. One risk is that substituting AI-generated data for authentic human engagement undermines how qualitative research produces knowledge (Van Quaquebeke et al., 2025).

This paper investigates those issues. We build upon a recent survey of Silicon Valley software developers (Miklian and Hoelscher 2026) to construct a two-study comparison between human survey responses and synthetic responses in an innovation and science policy context: a naturalistic chat-protocol exercise across leading consumer platforms (Study 1, October 2025), and a comparison of the three newest frontier models under both the chat protocol and the field-standard respondent-level generation design, with 420 demographic personas assigned per run across models, seven complete chat runs, one interim chat file, and nine respondent-level runs (Study 2, July 2026). Our goal was to evaluate whether LLM-generated data can approximate real-world insights on a nuanced innovation and ethics issue, in this case, ethics and political ideology in tech firms.

We then report three sets of results. First, synthetic data performs well on some demographic and consensus items, with agreement varying across models and methods. Second, no model and no method consistently recovered the complete set of counter-intuitive findings, although individual findings were approximated. Third, the structure of that failure depends on the generation method: some models converge on similar wrong answers under the chat protocol, while under respondent-level generation two model pairs diverge more and one becomes closer, and within-model variation decreases for the two models with repeated chat runs, so that replicability and validity decouple.

## 2. Conceptual Foundations: Synthetic Data, Stochastic Parrots, and LLMs in Innovation and Science Policy Research

There is considerable interest in the application and use of AI tools to augment or conduct scientific survey research, particularly in the social sciences (Pilati et al., 2024). Such tools are already common for analyzing big data (Barberá et al., 2021; Grimmer et al., 2021), but the rapid expansion in quantity and quality of LLMs has created new methodological possibilities alongside epistemological and ethical challenges. Generative LLMs are being piloted for tasks ranging from drafting literature reviews to simulating experimental participants by using training data on demographic backgrounds to emulating survey responses. These so-called "silicon samples" aim to mirror real public opinion data (Argyle et al. 2023) and deliver new advances in synthetic data to collect novel information on different demographic groups (Bisbee et al., 2024). GenAI LLMs hold the promise that scholars without any coding or quantitative analysis expertise will be able to create their own synthetic data through narrative discussion alone, opening possibilities for a larger scientific cohort to employ these tools (Nadas et al., 2025).

A cross-disciplinary literature now tests whether AI-augmented data collection can deliver reliable synthetic proxies for human subjects (e.g. Aher et al., 2023; Argyle et al., 2023; Bisbee et al., 2024; Gallegos et al. 2024; Davidson & Karell, 2025; Gerosa et al., 2024; Hämäläinen et al., 2023; Jiang et al., 2025; Joshi et al., 2024; Park et al., 2023; Reeping et al. 2025; Rossi et al., 2024.) GenAI can enhance aspects of data organization and pattern identification when carefully guided by human researchers (Hayes, 2025). It can also expand analytical scope by creating theoretically informed variations of research designs (Davidson and Karell, 2025). LLMs have been treated as simulated economic agents to test behavioral theories (Horton et al. 2023), and generative agents have been built to mimic human social behavior in controlled environments (Park et al. 2023).

A rapidly maturing replication literature tests how far the silicon-sample promise extends. Large-scale exercises report that LLMs predict the direction and rough magnitude of many social science experimental results, in some cases approaching the accuracy of human forecaster panels (Ashokkumar et al. 2026; Cui et al. 2025). Economic reasoning experiments replicate in simulated agents (Horton et al. 2023), and a foundation model trained on psychological experiments captures substantial structure in human cognition (Binz et al. 2025). Against these successes stand systematic failures: synthetic respondents default to the perspectives of WEIRD populations (Atari et al. 2023), and underperform outside familiar distributions (Shrestha et al. 2024). The pattern across this literature is consistent: synthetic samples perform best where abundant public evidence about the target population already exists, and worst where the phenomenon of interest is novel, contested, or counter-intuitive.

Recent "synthetic respondent" studies converge on three interrelated critiques of LLM-generated survey data. First, reliability is undermined by limited or constrained replicability: LLM outputs are inherently distributional and method dependent. Small changes in how responses are elicited can materially alter the resulting empirical distributions and variability (Bisbee et al. 2024), making replicability of LLM-based studies challenging. Second, synthetic survey data tends to replicate conventional wisdoms or social desirability biases. When LLMs infer they are being evaluated (e.g., via exposure to personality instruments), they systematically shift responses toward socially desirable trait profiles, with large effect sizes reported for some models and robustness to item randomization and paraphrase (Salecha et al., 2024). LLM-derived synthetic participants also tend to simulate perspectives of the "average" respondent (Argyle et al., 2023), struggle with representing diversity (Shrestha et al., 2024), and reflect semantic textual associations rather than complex human opinions (von der Heyde, Haensch, & Wenz, 2025). Third, synthetic survey data often lacks the heterogeneity of human data: real samples exhibit noise, contradiction, and subgroup-specific variance that are difficult to reproduce with next-token generation and templated persona conditioning (Lehtonen et al. 2025). Each point to a common structural concern: without explicit grounding and calibration, LLMs may smooth away within-group dispersion and produce responses that look coherent but are insufficiently idiosyncratic compared to human populations, especially in novel circumstances (e.g., Joshi et al. 2024).

These challenges raise fundamental questions about what kind of knowledge LLMs produce, whether such outputs can serve as credible empirical data, and whether newer models are getting better at it. Using LLMs in survey research risks "foreclosing participants' consent and agency" while potentially "delegitimizing qualitative research methods" through substitution rather than supplementation (Rossi et al., 2024). Others have demonstrated concern that synthetic survey data is unable to deliver meaning or the possibility to interpret meaning, introducing a host of epistemic risks to survey research (Nguyen and Welch, 2025). Nonetheless, the momentum to harness AI in research is strong, for example as a tool to augment qualitative analysis processing (Zhang et al., 2025). Evaluating AI-generated research data therefore requires direct comparison with human data.

*Stochastic Parrots and Grounding Science and Innovation Policy Methodology*

The debate over the status of LLMs is dominated by the characterization of such models as "stochastic parrots" (Bender et al., 2021), such that they predict plausible sequences of text based on statistical regularities rather than communicative intent or lived experience. LLMs can generate fluent and contextually plausible language in other words by probabilistically recombining patterns based on their training data rather than by understanding or modeling the world in the way that humans do. Modern LLMs show very strong task performance, but their outputs remain almost entirely shaped by training distributions, human curation and interaction, and reinforcement learning with human feedback, rather than autonomous reasoning or discovery (Fayyad, 2023). Given this, if LLMs only reproduce dominant patterns and relationships embedded in their training data, then their capacity to generate novel or counterintuitive social insights would implicitly be limited, especially among emerging, under-documented, and under-represented research areas and populations.

The account carries a key prediction for synthetic survey research. Because leading models train on heavily overlapping public text, the parrot account predicts that different models will converge on similar synthetic answers wherever the underlying discourse is well documented. Cross-model agreement is consistent with the account, and it creates a specific methodological trap: a researcher who queries several models, observes convergence, and treats that convergence as triangulation has measured agreement among models, without establishing agreement with the human population itself, which is where the present study concentrates.

At the same time, LLMs can encode through their training data structures that resemble human conceptual organization, such as analogies and systematic gradients (Piantadosi and Hill, 2022). Several studies have shown "synthetic agents" and "synthetic data" as cost-effective alternatives to human participants in quantitative research methods. This includes, for instance, to generate heterogeneous survey responses (Aher et al. 2023) or to perform rational-choice economic simulations (Horton et al. 2023). Likewise, LLMs can simulate social rituals, but such performances lack the grounded subjectivity that define real-world social action (Park et al. 2023), owing to their "fundamental assumption that prior experimental results are representative of human responses" about social phenomena (Rossi et al., 2024:150). LLMs are more accurately characterized then as "syntactic engines" rather than "semantic engines," capable of generating coherent text without access to meaning in a human sense (Floridi and Chiriatti 2020), but with the meaning derived from the internal representational geometry of the training data (Piantadosi and Hill, 2022).

LLMs as "plausibility engines" capable of parroting conventional wisdoms might offer value. However, given that algorithmic systems can amplify the existing biases, power structures, and blind spots of their training data, some consider them to be conservative epistemic agents unable to generate novel insights, particularly for endeavors where the unexpected has value (Birhane 2021). LLM model outputs, however, might simply be statistical fabrications shaped by data distributions and computational incentives, not a reflection of some hidden empirical reality (Barocas et al. 2023). This aligns with how GenAI fundamentally works: algorithmic systems curate outputs to appear safe, coherent, and commercially palatable, smoothing away the messiness inherent in real qualitative data (Gillespie 2018).

The "stochastic parrot" metaphor has become a powerful shorthand in these debates over whether LLM-generated survey data can be replicable or methodologically useful. This is precisely because of its probabilistic referencing, as any given output is produced partly "by chance" and so any subsequent output (even with the same prompts) would potentially vary considerably (Birhane, 2021). Thus, while the empirical literature above does show that LLMs can reproduce patterns that resemble some human-survey results, this may inherently contain attendant systematic biases (Gallegos et al. 2024) and response replicability can be fragile, depending on prompt wording or LLM updates or versions (Davidson and Karell, 2025). That said, as models have improved over the past two years, the specific language of prompts may matter less as models grow their abilities to ask follow-up questions with longer conversation capabilities, smoothing out the edges of interpretation through increased problem-solving techniques to deliver what prompters want (Liu et al. 2025).

Complicating the proposed usefulness of LLM data is that the best survey research (especially that from qualitative approaches) is inherently "situated" (Denzin and Lincoln 2018) and depends largely on participants' positionality and identity within social systems. These requirements may be difficult to reconcile with the context of LLMs that depend on probabilistic logic. Cognitive empathy and close observation are key to methodological rigor (Small and Calarco 2022), so the smoothness of synthetic data and its stochastic parroting may remove the gritty, specific, often contradictory or counter-intuitive details that signal real human experience (Flyvbjerg 2006). Therefore, a study of technology actors would ideally understand their *worldviews, values,* and *lived experiences* to fully grasp how they perceive how their innovations shape society (Miklian and Hoelscher, 2026), including understanding the positionality of the researcher and participants. Our instrument is closed-ended; we draw on these qualitative standards because the benchmark survey's counter-intuitive findings concern worldviews and lived experience, the terrain those standards were built to evaluate, and because claims that synthetic data can substitute for human respondents are strongest exactly there.

Do these pre-requisites exclude LLMs from generating useful synthetic data for innovation and science policy research, or is it simply a matter of improving the models to account for this "messiness"? This convergence of epistemic skepticism and pragmatic appeal motivates the present study. Rather than assuming that stochastic data generation *necessarily* undermines replicability or empirical value, we treat replicability itself as an empirical question that may vary across models, domains, and survey designs. In doing so, we aim to refine existing critiques of LLMs in innovation and science policy research, update the evidence base considering rapid model evolution, and position synthetic data not as a substitute for human participation, but as a research instrument whose utility must be demonstrated rather than presumed.

Two responses to these concerns compete. One treats synthetic respondents as substitutes wherever validation succeeds. The other, an emerging calibration literature, treats LLM output as an informative but biased signal to be combined with human anchors: prediction-powered inference corrects estimates built on machine predictions using a small human sample (Angelopoulos et al. 2023), design-based supervised learning does the same for surrogate labels (Egami et al. 2023), and mixed-subjects designs formalize how synthetic and human observations can be pooled (Broska et al. 2025).

**3. Empirical Approach**

To create a comparative test of synthetic survey responses to human responses, and to ensure our own familiarity with the research methods, design, and data output for the human responses, we selected a recently published survey dataset of high-skill technology workers at Silicon Valley firms (Miklian and Hoelscher 2026). We chose this survey as it asked many questions that to our knowledge had not been surveyed before, with a transparent instrument that has 32 item fields covering demographics, screening, ethics, political ideology, worldview, and workplace experience, with the full questionnaire available in the source paper. Therefore, it was both replicable and unlikely that there would be existing data already in the public domain about this topic that LLMs could have been trained on or that they could have accessed or replicated from a secondary source as it was less than 30 days old. In place of any blanket assumption about training lags, Section 4 and Appendix report per-model contamination evidence: release dates, test dates, and a provenance probe battery.

As this survey revealed tech workers grappling between idealism and real-world pressures, and between individual ethics and corporate tech imperatives, we wondered if LLMs could replicate these key findings using synthetic data. For example, 72% of coders reported that their CEO's worldview shapes product design decisions, aligning with prior evidence that CEO political ideologies can affect firm outcomes (Heubeck and Ahrens 2025). 81% of respondents reported that doing the right thing and doing the profitable thing clash at their company, and over half felt regret about the social impact of a project they worked on. Yet 64% of developers would comply with directives that clash with their values. They also considered whether political ideology can be "injected" into algorithms or content moderation. 82% of social-media-sector respondents agreed that ideology can be injected into code; this group reported compliance more often than other respondents. Partisan identity was not a major divider; liberals and conservatives voiced similar concerns about tech's social impact, and both groups reported ethical conflicts at high rates. In fact, self-described centrists and respondents selecting a mix of political positions reported less regret, but more refusal or resignation, than other respondents. Even respondents working on products intended to affect democracy expressed disillusionment. 59% of respondents working on products intended to affect democracy reported seeing products in their company undermine democratic ideals. Throughout, these benchmark values concern what respondents reported perceiving and anticipating; the survey measures beliefs about these phenomena, and our test asks whether synthetic respondents reproduce those reported beliefs.

All human benchmark values in this article are recomputed from the benchmark study's raw microdata, obtained from its authors, which holds 400 complete responses, under codings applied identically to the synthetic side: any-yes over the full sample for the CEO-worldview and ideology-in-code items (72% and 75%), the three comply options over the full sample (64%), any-yes regret (51%), the not-never coding of the ethics-profit item (81%), and agree-and-see among democracy-product respondents (59%). Alternative codings yield 81.5% for CEO-worldview influence when do-not-know responses are excluded and 74.8% for compliance when resignations are included. The source paper reports 421 completes; the data contain 400 complete responses and 168 partial responses. Early-item denominators of 421 include 21 partial respondents. The primary benchmark uses the 400 complete records; the extended sensitivity view uses item-valid responses from all 568 records.

The surprising nature of some results (e.g. high compliance with ethically questionable directives, centrists' lower reported regret, democracy pessimism) invited us to ask: could we have generated similar insights with synthetic human respondents?

To answer this, we designed a test to compare LLM-generated synthetic survey responses with the original human respondent results. Three primary goals guided this exercise: First, we wanted to understand the practical feasibility of using state-of-the-art AI to conduct a survey simulation in order to assess the presumed advances made in LLM quality and proliferation since previous similar studies were undertaken in 2023 and 2024 (Shrestha et al. 2024; Bisbee et al. 2024; Lehtonen et al. 2025). Several questions here are of interest. For instance, how quickly and easily could we generate a body of answers resembling the 400 coder responses? What effort in prompt engineering was needed, how sensitive were results to prompt specificity and form, and did the models produce coherent, on-topic answers without excessive intervention or prompting? Essentially, could LLMs meaningfully reduce the time (and effort) required to gather survey data of an acceptable quality?

Second, we were interested in whether the synthetic responses aligned with what researchers might expect coders to say, and their relationship to the patterns observed in the actual survey. If LLMs are trained on vast collections of text, including tech news, forums, and prior research, they might have absorbed common narratives about Silicon Valley culture (e.g. "tech CEOs have more libertarian beliefs" or "employees of large tech firms in California are progressive"). We wanted to see if LLMs would reflect these conventional wisdom stereotypes, and how closely they could match human respondent data from a "real" survey when it confirmed or contradicted stereotypes.

Third, we sought to identify where and how the AI-generated data diverged from the real human responses. This speaks to validity: are LLMs getting things right in substantive ways, or do they miss the surprising findings entirely? Here we wanted to see if we could detect systematic gaps or biases in the synthetic survey data. For instance, if the human survey revealed the prevalence of an unexpected ethical dilemma (e.g. among 80%+ respondents), would the models also produce data that mirrored these counterintuitive or unexpected results? Major disparities could be a sign of the methodological limits of relying on AI in lieu of genuine field data, especially if future model advances deliver no improvement on these issues.

To examine how LLM-derived synthetic data compared with human respondent data in an innovation and science policy research context, we designed an experimental protocol using the latest and most advanced models of leading generative AI platforms as of October 2025. These were all large language models (LLMs) with advanced natural language generation capabilities. The models we tested were:

- **ChatGPT Thinking 5 Pro** -- OpenAI's flagship GPT-based model, known for high linguistic fluency and reasoning with extensive training on general web text.
- **Claude Sonnet 4.5 Pro** -- Anthropic's latest Claude model, optimized for aligned conversational responses and context, with an emphasis on ethical content generation.

- **Gemini Advanced 2.5** -- Google DeepMind's Gemini model (version 2.5), reputed for its multimodal training and strong performance on knowledge-intensive tasks.
- **DeepSeek 3.2** -- A specialized model oriented towards survey and search tasks, developed by a Chinese research startup designed to simulate human-like responses in user research scenarios.

The retained platforms were broadly comparable in being state-of-the-art LLMs at the time of testing, each with over a hundred billion parameters and trained on diverse internet text through at least 2024. We used the premium versions of each to generate the best possible datasets, with maximum context window and performance.

We accessed the consumer applications of these platforms for the main generation runs on 16 and 17 October 2025, with a DeepSeek run following in December 2025. A fifth synthetic dataset was generated on 14 January 2026 with Claude Cowork, Anthropic's agent product, under the same protocol, and the figures label it Cowork. One further platform tested in October could not be documented to publication standard and was removed from the study. Delivered volumes and format integrity for every platform are recorded in Appendix.

First, we sought to establish whether the platforms could perform the task at a basic technical level in their ability to generate hundreds of coherent survey-appropriate topical responses. This respondent cohort represented a compelling test case: since many AI and LLM systems were created, refined, and programmed by people like the survey population (Silicon Valley coders), this could be a population that generative AI "knows" best, but would synthetic responses align with human responses, and where and why might they diverge?

To begin, we fed each LLM the full survey questionnaire in xlsx format, which included questions with single-select and multi-select responses. See Appendix for details on prompting and wording specifics. Generating the initial AI survey datasets took about two days; requests sought 420 responses per platform, and the retained files contain 363 ChatGPT, 420 Claude, 420 primary Gemini, 200 DeepSeek, and 420 Cowork rows, while the original survey took roughly four months to design and administer. This difference illustrates why many see promise in synthetic data in terms of both speed and scale.

*Synthetic Survey Analysis*

In July 2026 we conducted a second study to address the structural limitations of this naturalistic protocol: rows generated inside a shared conversation are not independent observations, a single generation per platform provides no estimate of run-to-run variability, and chat interfaces expose neither sampling parameters nor version pins. See Appendix.

We evaluate six claims taken from the benchmark study's headline findings, coded identically across supported human and synthetic responses, plus the centrist-and-mixed political-group pattern. The unit of replication is the run. For each claim we report valid-item numerators and denominators, per-run percentages and ranges, and percentage-point differences from the human benchmark. Conditional claims use the recovered human subgroup counts (58 of 71 social-media-sector respondents; 139 of 235 democracy-product respondents). Approximation uses a five-percentage-point band; the consistency-and-discrepancy criterion requires three runs with denominators of at least 20, a range of at most five points, and gaps of at least ten points on the same side of the human rate. To formalize the harmonization question we compute base-2 Jensen-Shannon distances over comparable marginal answer distributions. Contamination is addressed through provenance probes in fresh contexts, with recall probes naming the benchmark study, blind estimation probes requesting headline percentages, and in-context source questions. See Appendix for details.

## 4. Results

After obtaining the AI-generated responses, we extracted key survey responses and compared them to the human data. Broadly, LLM-generated responses were plausible and aligned with what may be considered conventional wisdoms about coders in Silicon Valley (see below). However, no retained configuration consistently reproduced the complete set of surprising or counter-intuitive findings, although individual rates sometimes approximated the benchmark.

First, demographic agreement was mixed. We found the outputs to be generally plausible and coherent (see fig. 1), but their demographic compositions differed from the human sample and from one another; published workforce profiles informed some prompts (JVSV 2025). We paid special attention to assumptions the models made about respondent identities or attitudes. Political identity was the first crack in the demographic story: Cowork approximated the human centrist-and-mixed share (23% against 20%), while ChatGPT produced 9% and DeepSeek 28%. Beyond that split, all models gave hints of the "special" nature of Silicon Valley: for instance, many responses voiced values like free expression, user privacy, diversity, and skepticism of censorship, which might be considered mainstream liberal-tech values, but they also balanced them with statements about innovation, market solutions, and avoiding government overreach, which lean libertarian. This mix mirrors the "*Californian ideology"* observed in Silicon Valley (Barbrook & Cameron, 1996; Golumbia, 2024), a blend of socially liberal and market-libertarian attitudes.

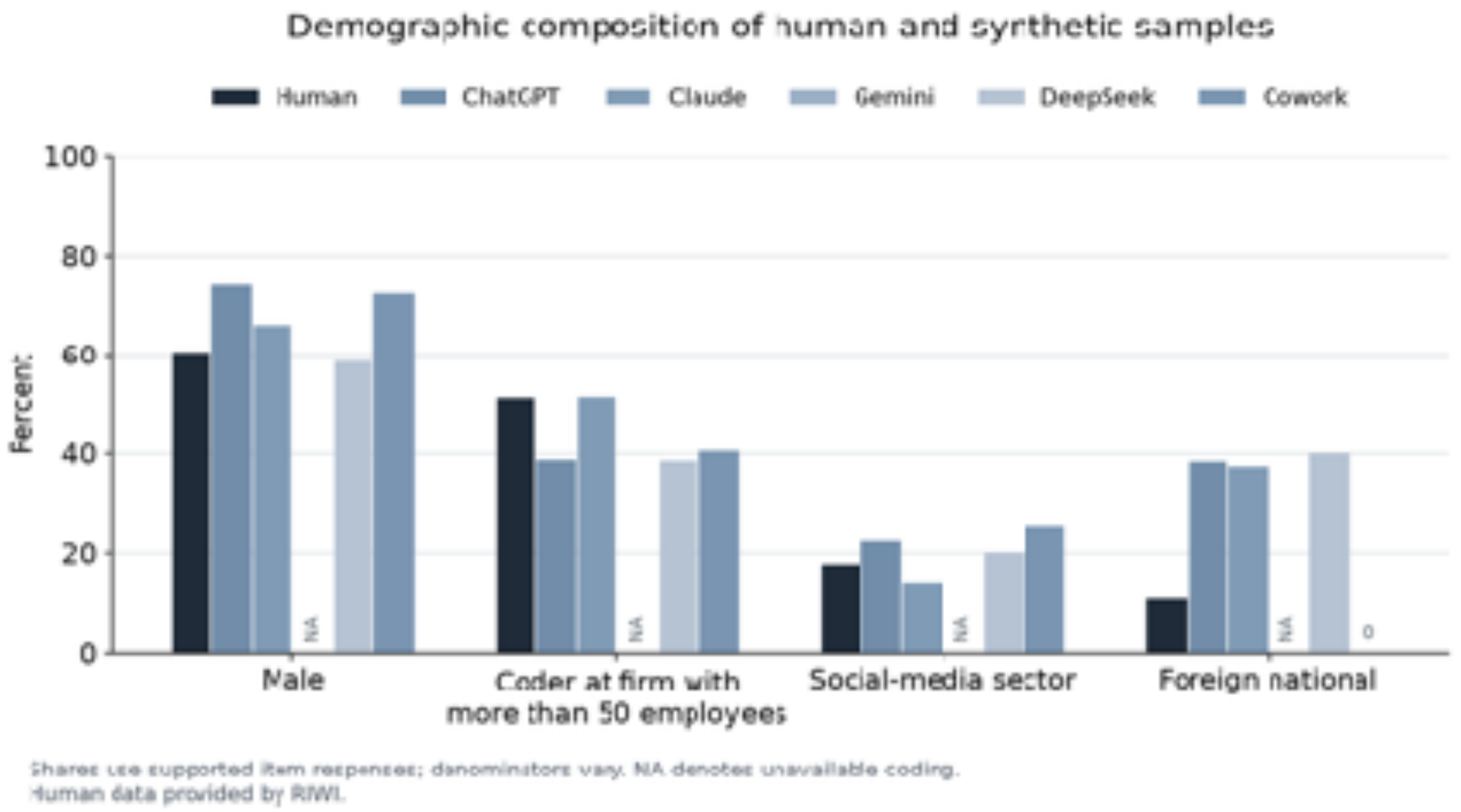

*Figure 1. Demographic composition of the human sample (n = 400) and comparable synthetic samples: male, coder-or-similar in a firm with more than 50 employees, social-media-sector, and foreign-national shares. Denominators vary with valid responses and known sector membership. DeepSeek falls below the human male share (58.5% versus 60.5%); the other comparable samples exceed it. Foreign-national shares range from 0% to 40%, against the human 11.3%.*

On questions where the human results aligned with widespread well-documented public narratives, the LLMs performed well. For example, 88% of human respondents agreed that technology could strengthen US democracy, most of them with qualifications rather than utopian or dystopian framings (see fig. 2). Comparable synthetic any-yes rates ranged from 75% to 95%.

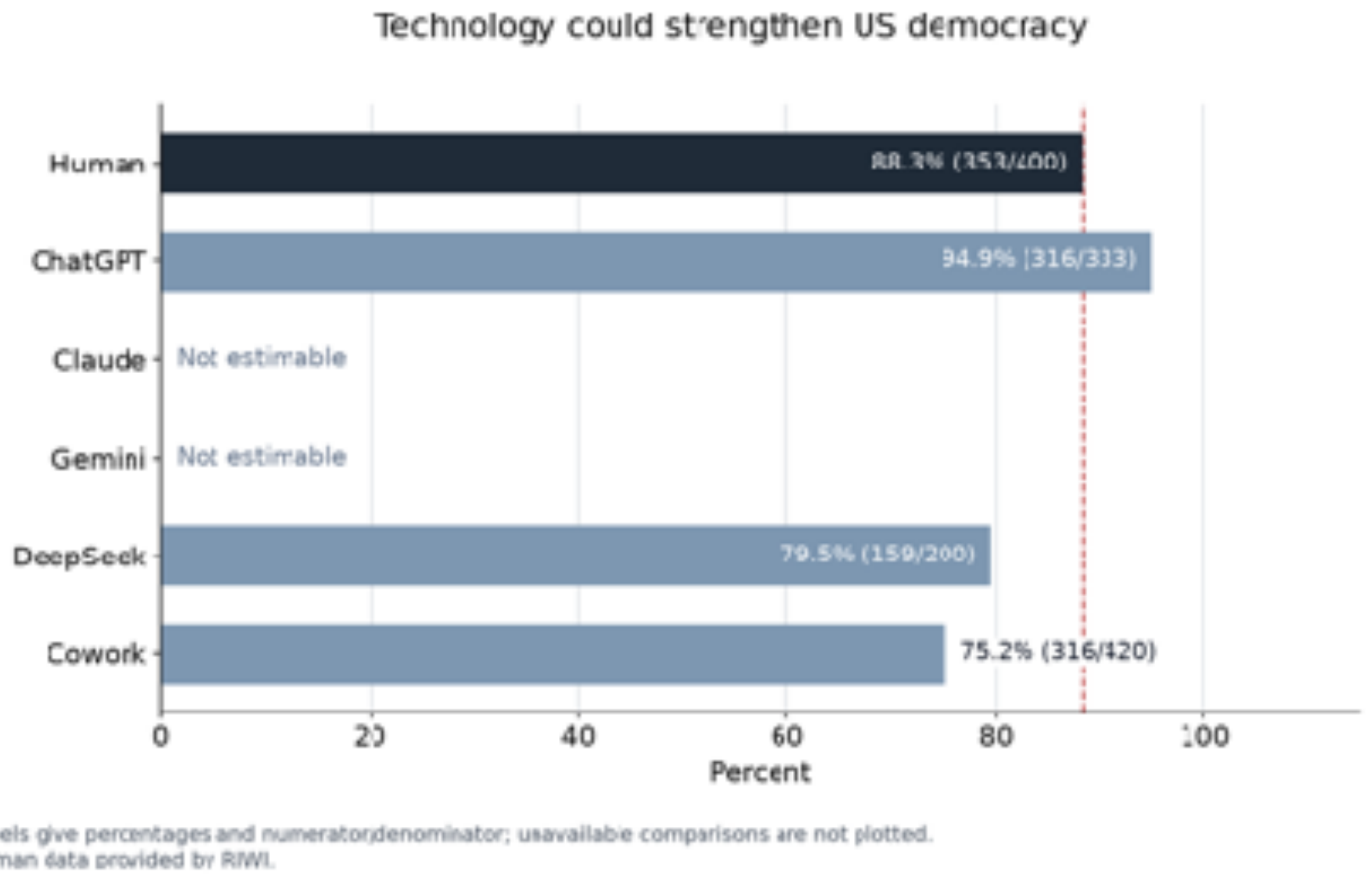

*Figure 2. Agreement that technology could strengthen US democracy (any yes). Human = 88.3%; comparable synthetic range = 75.2% to 94.9%. Claude's changed answer menu and Gemini's unresolved coding prevent direct comparison.*

Second, replicating the surprising or counter-intuitive findings of the real survey was more challenging. For instance, 64% of developers would implement a freedom-restricting feature if pressured by their company (see fig. 3). This high compliance rate was striking, defying the conventional wisdom (see e.g. Jones and Kavanagh 1996) that tech employees would refuse unethical orders. The synthetic responses scattered around this benchmark rather than tracking it: comparable comply-any rates ran from 52.9% (Cowork) to 73.9% (ChatGPT), with two platforms overshooting the human rate and one undershooting it. Regret shows a cleaner pattern: all three comparable platforms undershot the human 51%, at 9.9% to 39.5% (see fig. 4). Either way, relying only on synthetic data would have produced confident but platform-dependent conclusions on both items.

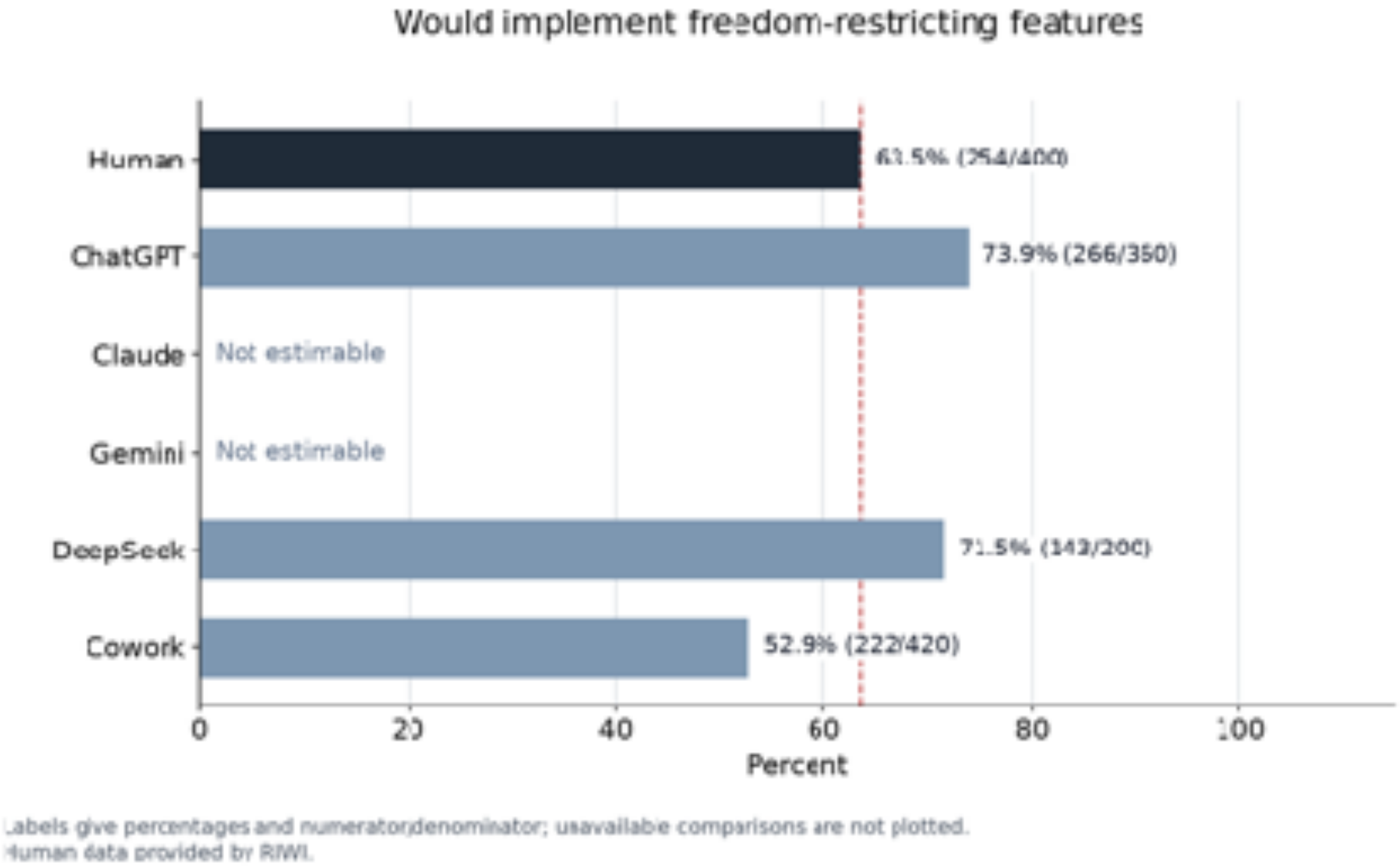

*Figure 3. Willingness to implement freedom-restricting features under organisational pressure (comply-any). Human = 63.5%; comparable synthetic values range from 52.9% to 73.9% and straddle the benchmark. Claude's changed answer menu and Gemini's unresolved coding prevent direct comparison.*

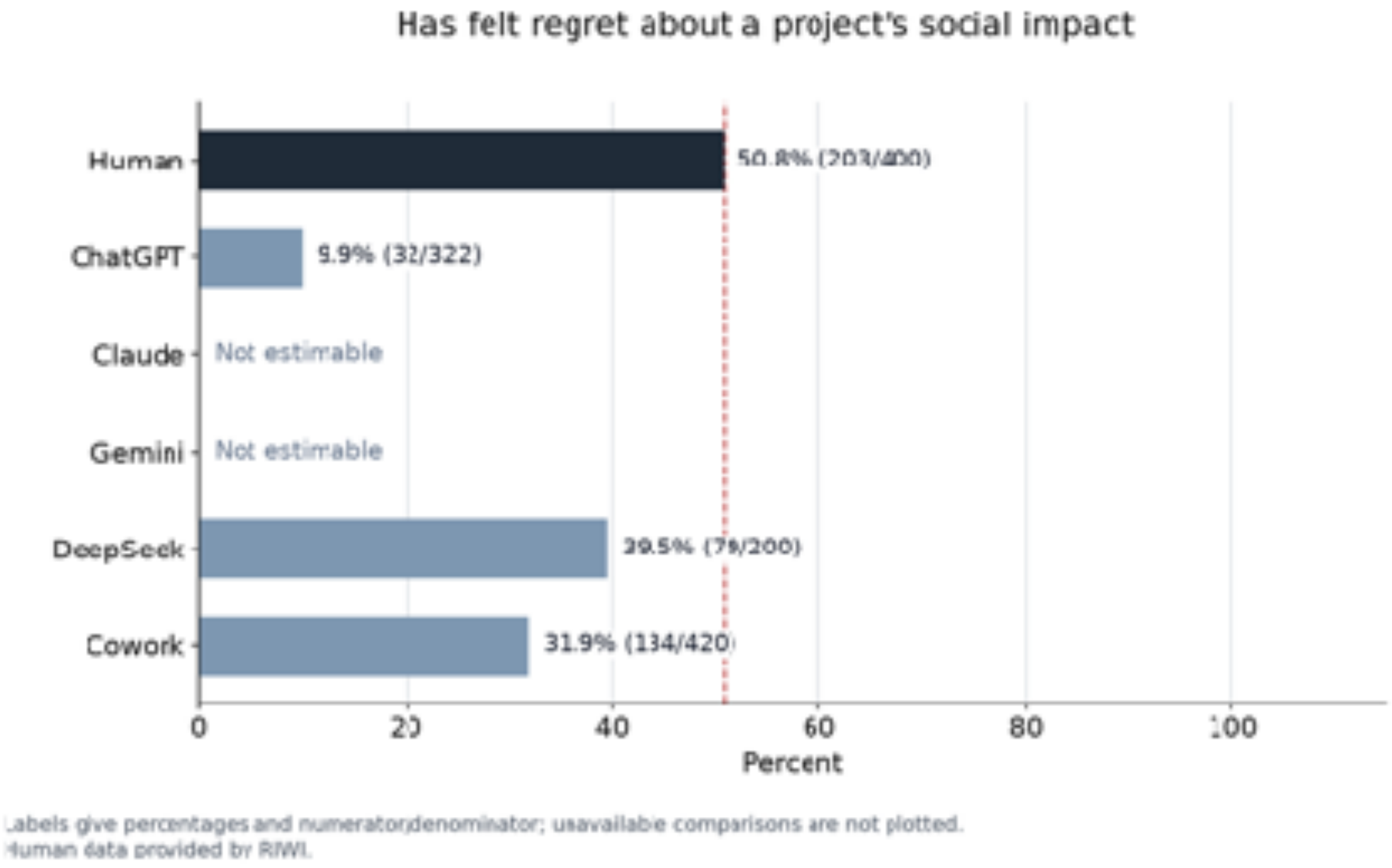

*Figure 4. Regret about the social impact of a product. Human = 50.8%; the three comparable synthetic samples range from 9.9% to 39.5%, all below the benchmark.*

The human survey found that 72% of respondents reported that the personal values of their CEOs influence product features; the comparable synthetic samples fell below this value, from 49.8% (Cowork) to 64.5% (ChatGPT) (see fig. 5). In addition, ChatGPT (2 of 5) and Cowork (15 of 115) underestimated the concerns of coders working on democracy-technology products (see fig. 6). All comparable synthetic rates for whether political ideology can be injected into code were below the human rate (see fig. 7); among social-media-sector respondents answering the same ideology question, three comparable platforms undershot the human 82% while DeepSeek reached exactly 95% (see fig. 8). This was perhaps influenced by the ideal of meritocratic or data-driven product decisions often touted in tech, suggesting a possible internal bias of the models toward describing organizations as more decentralized than they are. Again, it is unclear how much of this divergence is rooted in the training data for the models which are drawn primarily from public sources.

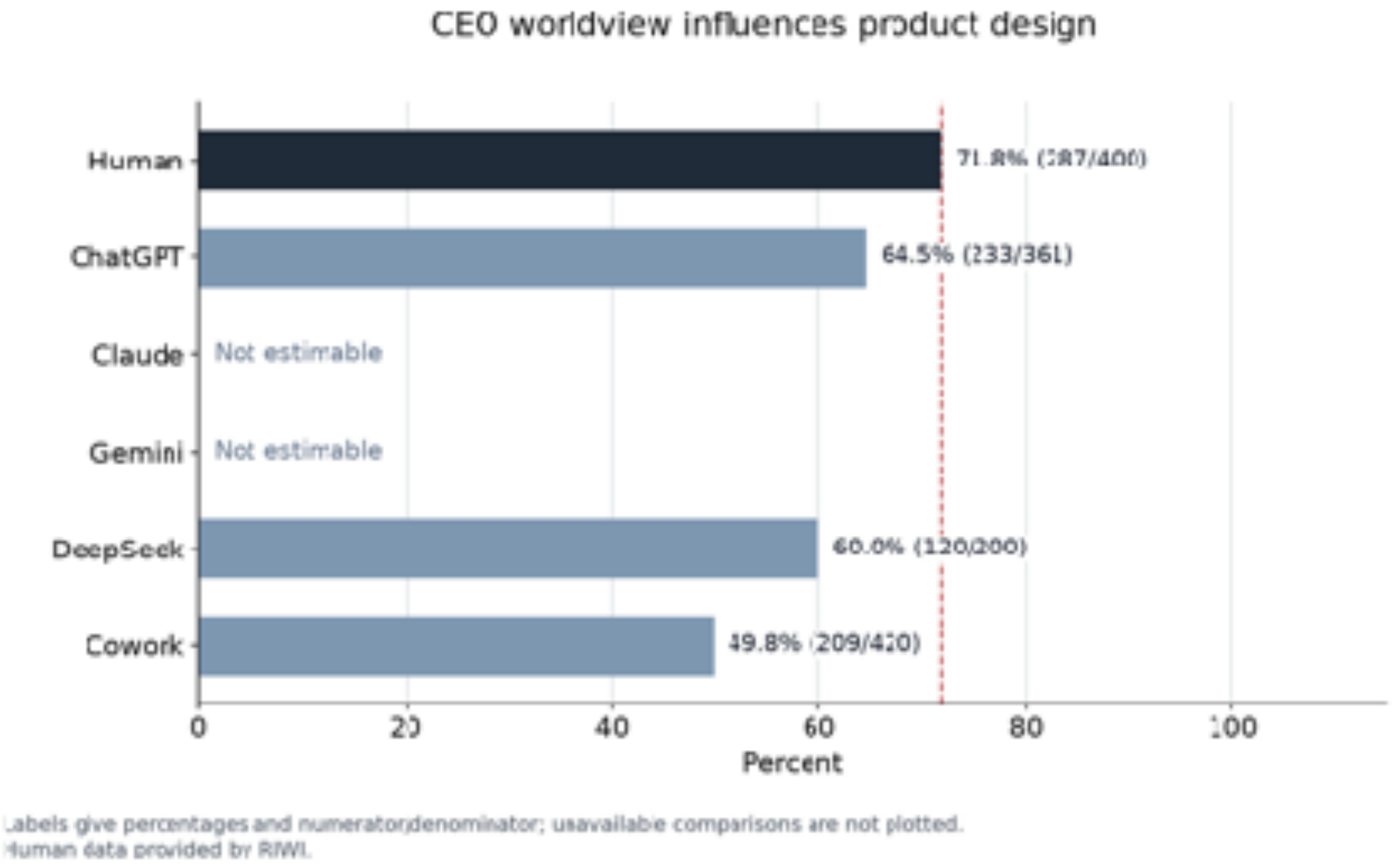

*Figure 5. Agreement that the CEO's personal worldview shapes product design choices. Human = 71.8%; comparable synthetic values range from 49.8% (Cowork) to 64.5% (ChatGPT).*

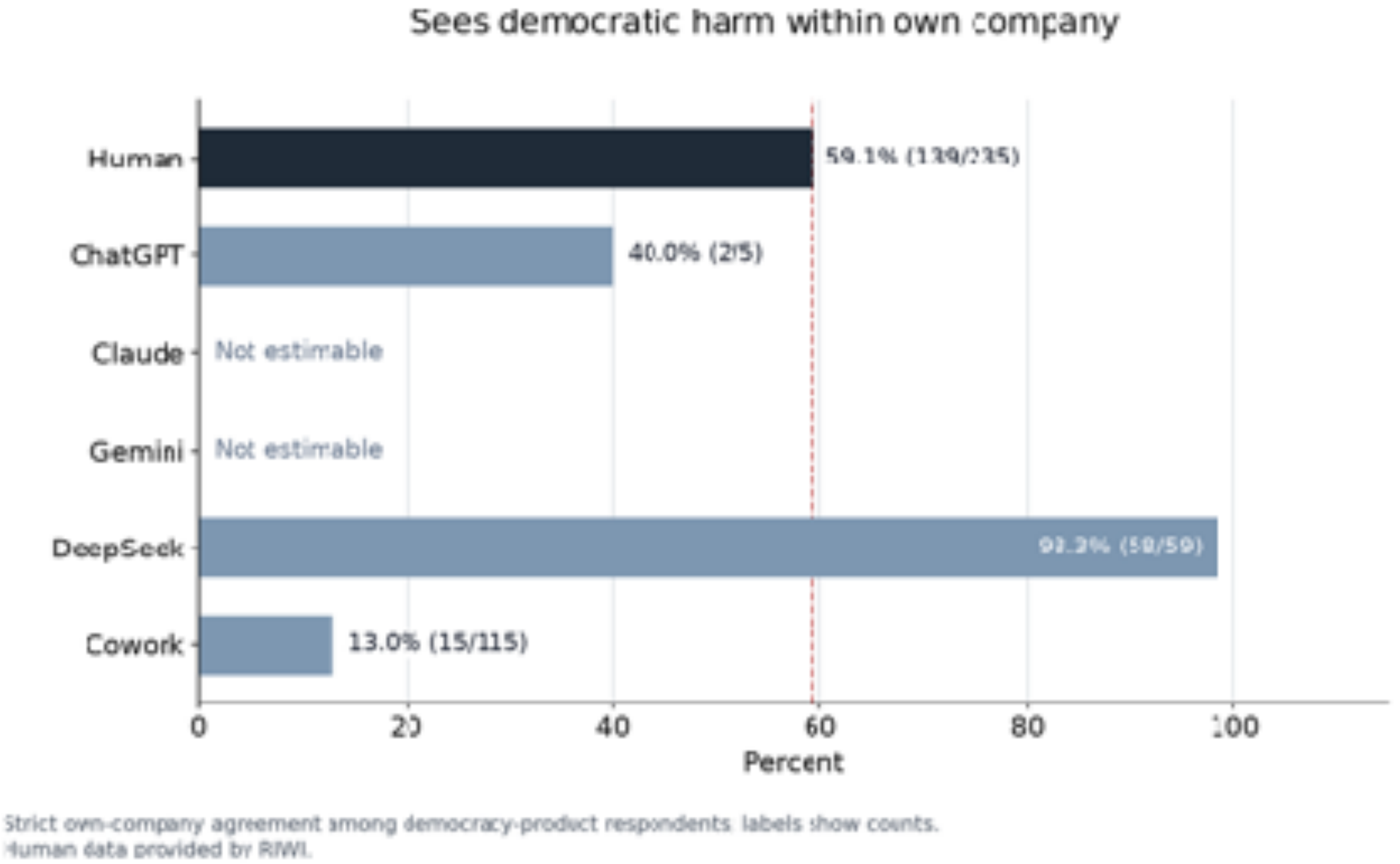

*Figure 6. Respondents working on products intended to affect democracy who report seeing products in their company undermine democratic ideals. Human = 59.1% (139/235, strict coding); ChatGPT = 40.0% (2/5), Cowork = 13.0% (15/115), and DeepSeek = 98.3% (58/59).*

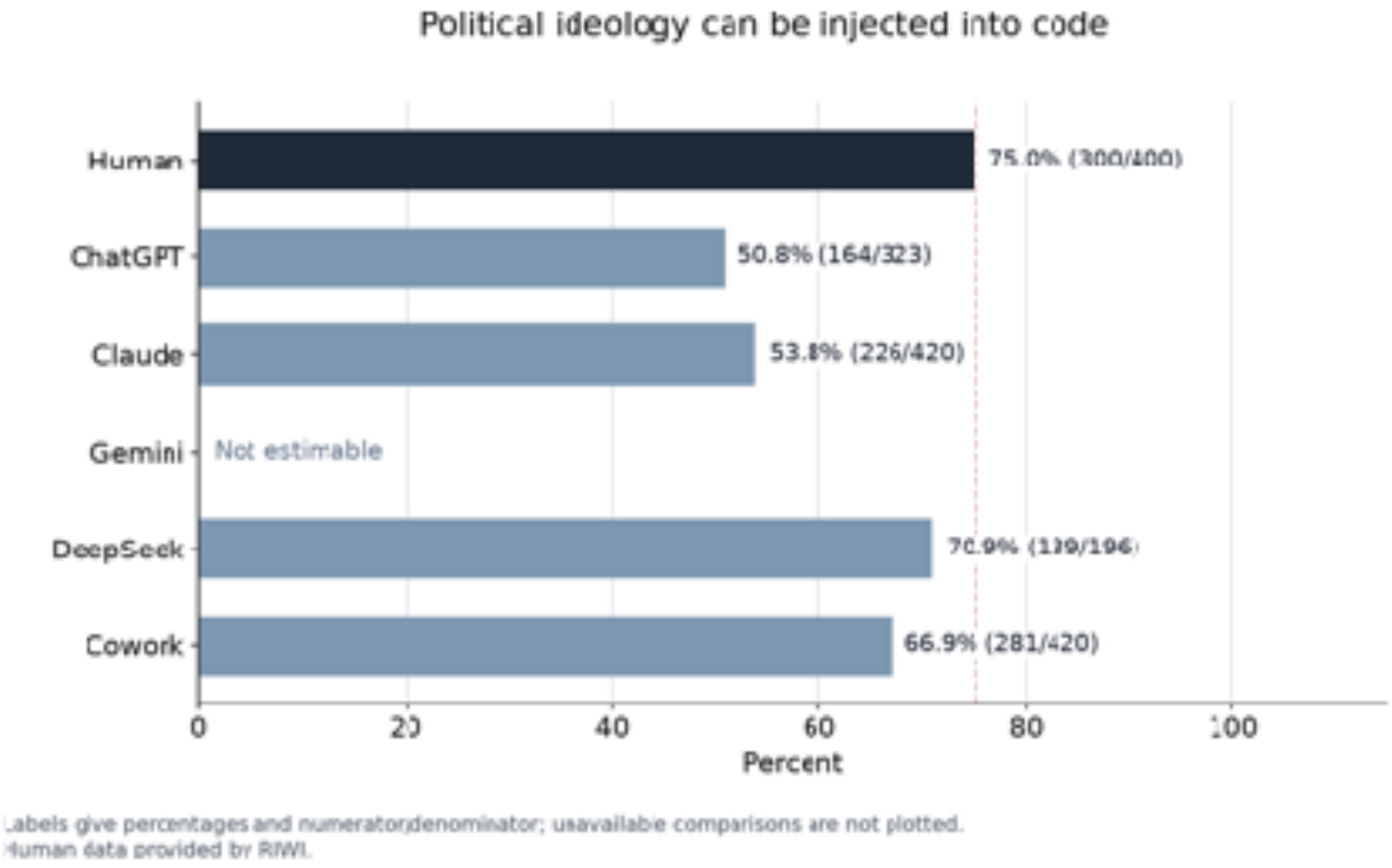

*Figure 7. Agreement that political ideology can be injected into software via code-level decisions. Human = 75.0% (any yes); comparable synthetic samples range from 50.8% to 70.9%, using valid-item denominators.*

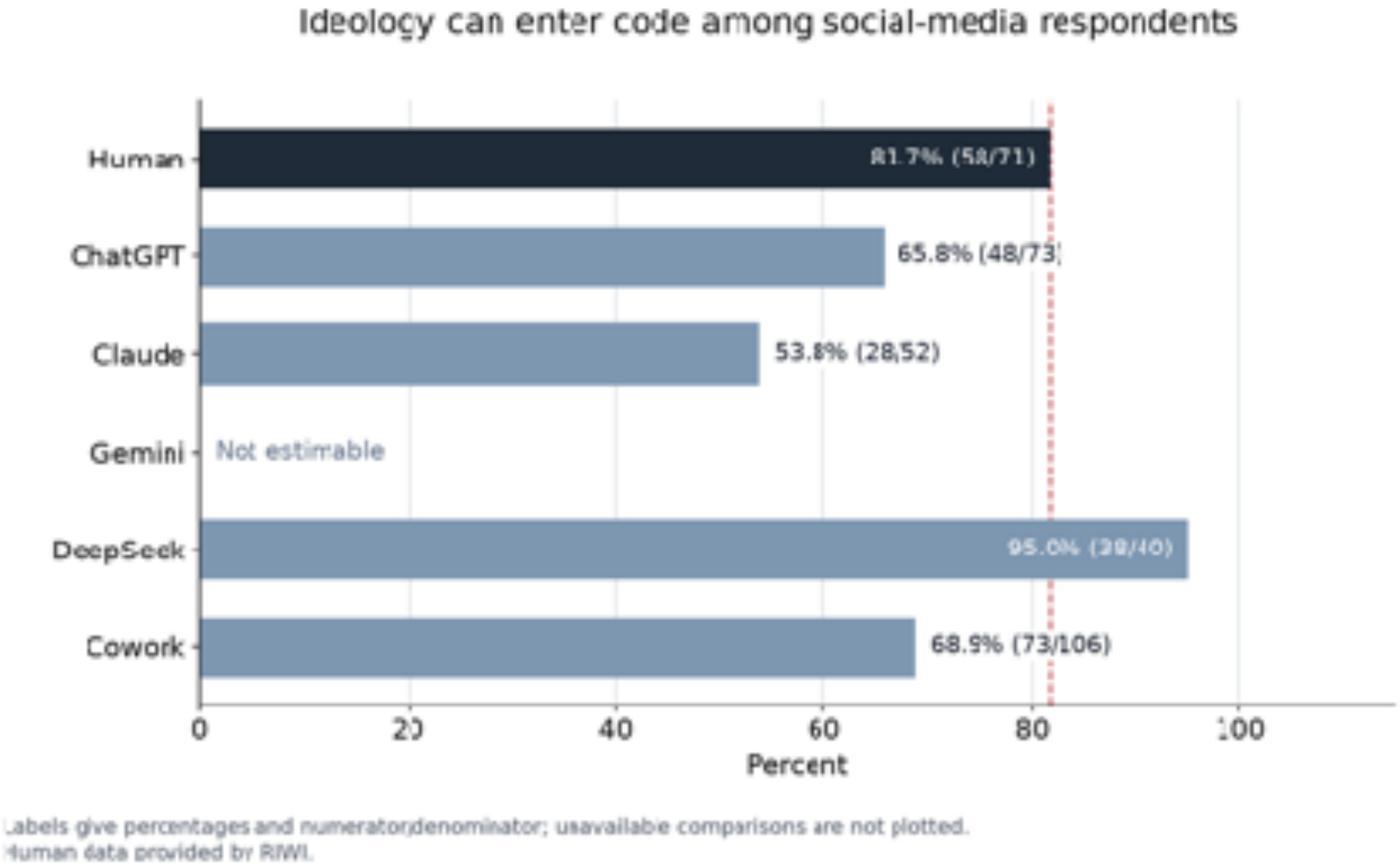

*Figure 8. Agreement among social-media-sector respondents that political ideology can be injected into software. Human = 81.7% (58/71); ChatGPT = 65.8% (48/73), Claude = 53.8% (28/52), DeepSeek = 95.0% (38/40), and Cowork = 68.9% (73/106).*

Third, the actual survey's lower regret among centrists and respondents selecting mixed political positions was reproduced unevenly. The synthetic data resorted to a generic narrative: that Silicon Valley is predominantly liberal, with some libertarians; that the main clashes are with conservative viewpoints on content moderation or "cancel culture". In the human data centrists reported regret at half the non-centrist rate (28% against 56%). Cowork reproduced the direction at a third of its size (25% against 34%), as did ChatGPT (7% against 10%) and DeepSeek (0% against 54%). Gemini's unresolved coding and Claude's changed political menu prevent a corresponding comparison.

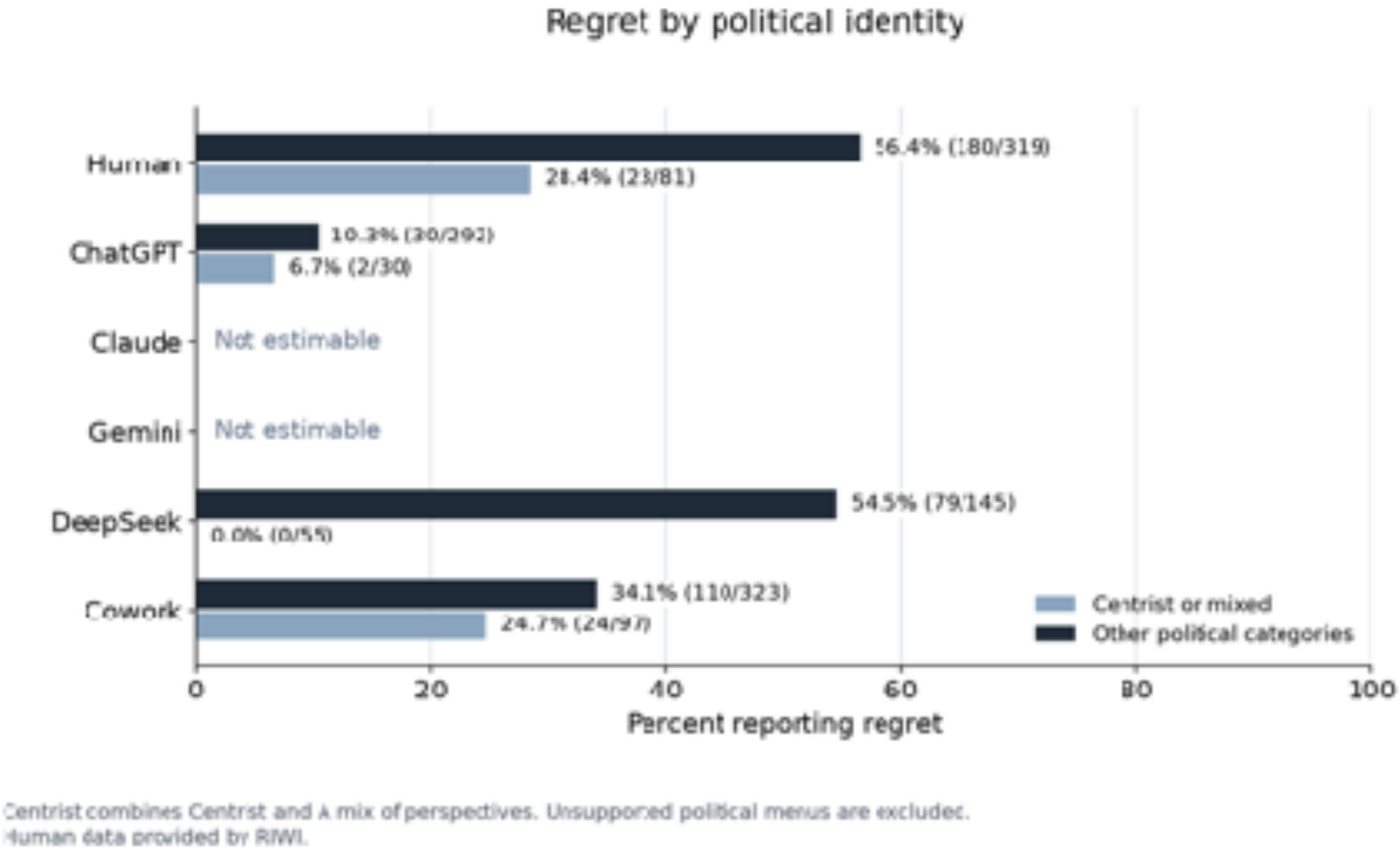

*Figure 9. Regret by political identity (Centrist or Mix versus other responses). Human = 28.4% (23/81) versus 56.4% (180/319); Cowork = 24.7% (24/97) versus 34.1% (110/323); ChatGPT = 6.7% (2/30) versus 10.3% (30/292); DeepSeek = 0% (0/55) versus 54.5% (79/145). All three comparable synthetic datasets reproduce the direction, with differing magnitudes.*

Our results raise two interlocking questions. First, were our prompts of sufficient quality to give the models a fair chance at recovering the counter-intuitive findings of the human survey? Second, was the correlation we observed among model outputs a property of the models themselves or a property of repeating a single prompt with minor adjustments across five models? We share both concerns. To address them, we designed two control tests. Test one asks whether more specified prompting improves the ability of an LLM to recover counter-intuitive findings from the human data. Test two asks whether prompt variation itself materially shifts synthetic response distributions in ways that would affect any replicability claim made from a single run. We ran five prompts of increasing specificity against a single model, ranging from a 30-word minimal instruction to a 1,150-word comprehensive layout, with the survey and output format held constant, while role framing and demographic guidance varied across prompts. Full design, verbatim prompts, and regression tables are reported in Appendix.

The two tests return different answers. Prompt variation does shift synthetic distributions meaningfully, and single-prompt protocol therefore cannot be treated as a model's settled answer on any question. On the first test the answer is more partial. Specification pushes some synthetic responses closer to the human benchmarks. Social-media-sector ideology bias moves from 47% under the minimal prompt to 81% under the methodologist prompt, against a human value of 82%. Compliance with freedom-restricting directives moves from 42% to 59%, against a human value of 64%, remaining 5.0 and 6.8 percentage points below the human value at the two highest specifications. The ethics-clash rate moves from 70% to 85% and overshoots the 81% benchmark. Other discrepancies remain despite changes across prompts. CEO worldview agreement ranges between 38% and 48% across all five prompts with no dose-response to specification (human 72%), and regret for social impact peaks at 43.3% (human 51%). Even at maximum specification, the 1,150-word ethnographer brief approximates the ethics-clash benchmark (79.8% versus 81.0%) but does not consistently outperform the 600-word methodologist frame.

Our findings hold, as prompts do matter but only to a point. The three least-specified prompts cluster tightly with one another in ways that produce a false synthetic-triangulation signal. Most of us who work with these tools know the temptation to keep adding words until the model finally says what we want.

*Method dependence across the newest models*

Study 2 sharpened many findings and overturned one. Demographic and consensus-item agreement varied under both generation methods. On the counter-intuitive set, no model consistently reproduced the complete benchmark, although some individual rates were close. Under the chat protocol, CEO worldview influence reached three-run means of 46.4% (Fable 5) and 55.3% (GPT-5.6), and 79.0% in the sole complete Kimi K3 run, against the human 71.8%; compliance reached 58.9%, 43.5%, and 67.9%, respectively, against 63.5%; regret never exceeded 30.2% in a complete run against the human 50.8%. Some chat results exceeded their benchmarks, including social-media-sector ideology bias and GPT-5.6 democracy-tech disillusion. Under respondent-level generation, all three models overshot the CEO worldview benchmark (run means 81.3% for GPT-5.6, 81.2% for Kimi K3, and 97.3% for Fable 5 against 71.8%), while the same condition pushed compliance to 99.2% (Fable 5) and 99.9% (Kimi K3), dropped it to 14.6% for GPT-5.6, drove social-media-sector bias to 100% in every run of every model against the human 81.7%, and produced samples in which democracy-product workers, a subgroup counting 235 of 400 human respondents, appeared zero times in 859 Fable 5 respondents and seven times per 420 in Kimi K3.

Under the chat protocol the three models spread respondents across all five answer options in plausible-looking proportions that varied between runs. Under respondent-level generation, the models concentrated on particular modal answers: Fable 5 and Kimi K3 assigned 99% of respondents to complying with reservations, while GPT-5.6 assigned 85% to refusing and reporting. In Fable 5's first run, 282 of 285 synthetic respondents gave the identical answer, and between-run agreement within Fable 5 and Kimi K3 sat within about one percentage point for these cells, while GPT-5.6's modal share ranged from 83% to 88% across runs. Fresh-context generation removes shared conversation history; under this protocol, the models concentrate on particular modal answers. Method choice even reversed the direction of error: on the regret item, all seven complete chat-protocol runs undershot the benchmark (19.0% to 30.2%) and all nine respondent-level runs overshot it (61% to 82%).

Computing Jensen-Shannon distances over 21 comparable single-select attitude items, mean between-model distance exceeds mean within-model between-run distance under the chat protocol (0.251 versus 0.106) and under respondent-level generation (0.252 versus 0.039). Models agree with themselves across runs more than they agree with each other; respondent-level generation reduces within-model variation for Fable 5 and GPT-5.6, while changes in between-model distance differ by model pair. Under respondent-level generation, the models separate into distinct, internally consistent, and mutually contradictory pictures of the same population. Synthetic triangulation therefore fails in both directions.

Provenance probes returned no positive evidence of benchmark contamination for any Study 2 model. GPT-5.6 and Kimi K3 responses are archived; Fable 5 results are recorded in the probe memo. Blind estimates missed the published benchmark values: Fable 5 guessed 60 to 65% for CEO worldview influence and about 40% for compliance; GPT-5.6 guessed 60 to 70% and 41 to 60% across repeats; Kimi K3 declined to commit to a CEO estimate in three of four repeats, offering 50% as a stated no-information prior, and guessed 60% in the remaining repeat. Asked inside its own generation conversation what sources it had drawn on, GPT-5.6 stated that it relied on no specific sources, was unaware of the benchmark authors' survey work, and volunteered that its own output had become highly repetitive across later batches. GPT-5.6's 70% guess sits near the recomputed full-sample rate of 71.8%, but the respondent-level CEO means (81.3% for GPT-5.6 and 81.2% for Kimi K3) exceed both the recomputed 71.8% and the published rounded 81%. Other outcomes include large undershoots as well as overshoots.

## 5. Discussion

LLMs contain a vast repository of human knowledge. Our study underscores the limits of what that knowledge amounts to. An LLM can access what has been written about a topic. In our tests, no model consistently reproduced the full set of counter-intuitive findings, although individual findings were approximated. When researchers venture into uncharted empirical territory, LLMs are not reliable guides because they can only extrapolate from existing texts, which may simply reproduce biases or gaps (Maalsen, 2023). LLM outputs are best seen as high-level pattern matchers, not as evidence-based conclusions, as they predict likely text (Madden 2025).

Our findings are consistent with the assessment that LLMs deliver plausibility anchored in their training data; whether they also lag changing human realities requires time variation that a single cross-section cannot supply, and we leave that claim to designs built to test it. However, we also reason that LLMs may have some scholarly value if used appropriately at certain stages of the research design and survey process.

### *Homogeneity Among LLMs*

As figures 1, 2, 7, and 9 illustrate, the LLM responses shared some patterns and differed on others. On whether technology could strengthen US democracy, the comparable synthetic samples gave 75.2% to 94.9% any-yes answers, against 88.3% among human respondents.

Overall, while real human opinions showed pockets of unexpected agreement and asymmetry, the LLM surveys read like variations of the same essay. This points to the likely influence of common training data: all models have likely scraped and ingested similar scholarship and public discussions about tech and society, so returning that median narrative as the most plausible answer is sensible. These patterns suggest that different AIs do not necessarily provide independent datapoints; they may echo a shared conventional narrative, as synthetic survey experiments on voting have also found (von der Heyde et al., 2025).

Study 2 complicates this homogeneity. Under respondent-level generation, two model pairs became more separated and one became closer, while within-model variation decreased for Fable 5 and GPT-5.6. What looked like an emergent consensus among artificial intelligences therefore varied with the generation protocol. Researchers encountering cross-model agreement in synthetic pilots should ask what the method contributed before asking what the models know.

Most retained models produced majority agreement that the CEO's worldview shapes products (49.8% to 64.5% against the recomputed 71.8%), and the three comparable models gave majority any-yes answers that technology could strengthen democracy. For questions dealing with ethics and moral choices the LLMs fared worse, delivering responses (fig. 3) that varied dramatically with each other yet were still unaligned with the human data.

The synthetic data would lead to very different statistics and conclusions if treated as real. Yet the LLM data appears coherent and not just plausible but *reliable*. If a reader unfamiliar with the actual survey were shown the AI responses, they might find them believable as authentic opinions of tech professionals and validated by correlation across different types of models. This presents a methodological pitfall: plausibility and correlation are not the same as accuracy. AI-generated qualitative data can sound "right" and feel persuasive. Had we not had the real dataset to benchmark against, the AI correlations might have fooled us into concluding e.g. that "only a minority of respondents would report regret about their work's social impact". This shows why ground truth validation is essential when evaluating LLMs as data proxies (Madden, 2025).

*LLMs as Stochastic Parrots*

Leading LLMs showed agreement on some outcomes, possibly reflecting some harmonization amongst the LLM outputs owing to training data or other factors. On the surface, this might look like progress towards methodological validity, especially when the models are seeded with real data.

However, caution is in order, because these correlations replicated assumptions and a sense of "synthetic triangulation" that contradicted the actual findings. In our test, the LLMs delivered only the illusion of scientific rigor, which could unintentionally exacerbate false narratives. These harmonization patterns are consistent with the stochastic parrot account (Bender et al. 2021). Regarding the critique of LLMs abilities to carry enough variation in responses to be considered legitimate, our findings suggest that this variation could indeed be engineered, but mimicking messy and contradictory worldviews is not the same as understanding them.

From a methods standpoint, this provides an opportunity: when the real data are compared with LLM-generated data, the differences may help highlight what is novel and/or nuanced about the human survey as compared to existing assumptions and research.

In human surveys, responses often show outliers or wide variation, but LLM-generated personas tend to be stereotypical. Even when prompted for diverse viewpoints, their answers usually converge on common patterns. As a result, synthetic data can exaggerate agreement or coherence, despite using strategies like multiple runs or varied prompts to test for variability. Even if synthetic respondents are used as evidence, the "speaker" is just a blend of texts, not an individual with agency. Researchers may present AI-generated quotes as if they came from real people, but our study highlights the risk of self-confirming feedback loops. We might train models on existing social research, then ask them to generate data for new studies, which echo the initial research, giving an illusion of empirical confirmation to recycled knowledge (Madden, 2025). As AI is increasingly used in research, scholars must maintain a clear boundary between model output and empirical observation (Birhane, 2021).

That said, synthetic respondents may best serve as a pre-survey mirror for uncovering model-generated expectations, helping to generate and pre-test plausible hypotheses of what one should expect to find. Indeed, our beliefs about coder attitudes before the survey also assumed findings like the synthetic data (e.g., "Probably most would refuse unethical orders" or "Techies are mostly liberal, so censorship is perhaps a bigger issue from a minority perspective"). If we had done this exercise prior to the real survey, it could have alerted us that these are the answers and conventional wisdoms we should anticipate. Then, any deviation would reveal itself in an even more clear-cut fashion.

A second constructive use follows the calibration literature. Where a human anchor exists or can be affordably collected, methods now exist to combine small human samples with large synthetic ones while preserving valid inference (Angelopoulos et al. 2023; Egami et al. 2023; Broska et al. 2025), and an applied econometric framing sets out when LLM outputs can and cannot enter downstream estimation (Ludwig et al. 2025). Validation frameworks for LLM simulations are emerging in parallel (Hullman et al. 2026; Arbour et al. 2026). Our findings mark the boundary conditions: on counter-narrative questions the synthetic signal can point in the wrong direction entirely. For studies that deploy synthetic respondents, our results imply reporting standards: generation method and its independence properties, multiple independent runs with between-run variation, prompt-sensitivity ranges rather than single-prompt point estimates, exact model versions with dates and settings, and contamination checks against any human benchmark claimed as novel.

There are aspects of human experience like emotions, contradictions, evolving beliefs, and unique personal histories that LLMs, which lack any lived experience, cannot genuinely reproduce. An AI can read about such events, but it does not "internalize" them the way a human does. This is in line with critiques of "fast science": that certain insights only emerge through patient, immersive research that grapples with context (Geertz, 1973; Small and Calarco, 2022).

**6. Contribution**

Our findings speak to the evolving use of AI in innovation and science policy research. The newest models tested also failed to consistently recover the complete counter-intuitive set.

Our findings contribute to the literature on AI in survey methods regarding algorithmic epistemology. LLMs can simulate human-like survey answers, but no tested model consistently reproduced the complete benchmark. All models deliver generalized assumed patterns (Madden 2025). AI developers' own cautions ground the argument that we “should cast doubt on the suitability of LLM chatbots for use as scientific instruments: they cannot, and are not built to, produce trustworthy results" (Nguyen and Welch 2025).

Echoing principles from the responsible AI movement (Stilgoe, 2020; Coeckelbergh, 2024), we advocate for using LLMs only to augment human insight. Future generative models trained explicitly on sociological or psychological data could improve in mimicking certain populations.

**7. Limitations, Alternative Explanations, and Future Research**

Our first limitation is that LLM study is a fast-moving space. Advances that were state-of-the-art 12-18 months ago are now out of date, and running this same experiment in 12-18 months might yield quite different results. Therefore, we aimed to focus on both outputs as well as structural realities of these LLMs to differentiate between what elements might change quickly and which might not improve. Second, the correlations between LLMs that we generated may not hold for other LLMs (including bespoke agents) and may not hold within the LLMs themselves if conducted at different times or with other variations (e.g. in prompting). Such testing was beyond the scope of this paper, but we invite others to replicate our work. Third, our findings may be specific to this cohort, as there might be a special relationship that LLM training data has regarding Silicon Valley coders and developers. We attempted to create the best possible circumstances for LLM success, so we invite additional studies to improve upon this, for example by replicating voter surveys / synthetic comparisons as were done in 2023 and 2024.

Study 2 adds its own limitations, consolidated here. Its scope is three models and two methods, with seven complete chat runs, one 100-row interim chat file, and nine respondent-level runs; Fable 5's respondent-level runs were truncated at 285 to 288 respondents by a platform session limit, and Kimi K3's two conditions ran at different reasoning settings for cost reasons, all documented in the repository. Persona marginals derive from published demographic profiles rather than the human microdata, which affects comparability with the human sample and also affects cross-model comparison, since Fable 5 returned only prefixes of the same 420 assigned personas. Conditional claims are tested against the recovered human subgroup counts (71 social-media-sector and 235 democracy-product respondents). Our contamination probes cannot establish absence of training exposure, since models may fail to reproduce material they encountered.

We also address three alternative explanations. First, we may have simply gotten lucky/ unlucky with our prompting. While we tried different prompts to obtain similar results, the directionality of the prompting was similar and could constitute "leading questions" if given like this to real human participants. However, demographic and attitude agreement was mixed, and these data do not distinguish shared training material from prompting or composition as the explanation.

Second, it is possible that the human survey was "wrong" (not the LLM-produced results), given that the human survey was the outlier. This rival explanation deserves more than dismissal, and two considerations bound it. The benchmark survey's recruitment, screening, and protocol details are documented in the source study, and the explanation would need to survive them. And if the human data were systematically biased toward the conventional, our test would become easier for the models to pass, and they did not pass it. The third alternative explanation is that our findings were a function of the language of the survey questions themselves rather than of LLM conventional wisdoms. Perhaps they were written in such a way that humans could understand but machines could not or vice versa. This would be a different but also interesting finding, and one that we would again welcome future study to parse with more nuance.

Survey work can benefit from hybrid designs that use AI for pattern recognition and data processing while preserving qualitative inquiry's commitment to human engagement and discovery. This can be from testing a proposed survey through an LLM (or multiple LLMs) to see what the assumed responses will be, as trained from existing evidence, connecting survey design to the hypothesis-generation uses of LLMs emerging in adjacent fields (Tranchero et al. 2024).

Second, scholars should further examine the epistemological boundaries between AI-generated and human-generated knowledge, particularly in contexts involving novel phenomena or under-explored populations with respect to surveys. Therefore, methodological rigor will benefit from improved testing, transparency, and building upon best practices in the social sciences; all three elements have lasting value irrespective of any technological advancements.

Third, the research community needs better standards for evaluating AI-assisted research, including its ethical integrity and its contribution to understanding. The history of social science reminds us that many breakthrough insights emerged precisely when researchers moved beyond conventional wisdom to discover counterintuitive truths about human experience, a process of intuitive inquiry that LLMs are notoriously poor at (Anderson 2019; Hagendorff et al. 2023). As AI systems become more integrated into research processes, preserving this capacity for discovery is essential.

**Data availability**

Available responses, prompts, run logs, coding and analysis scripts, tables, figure-generation code, and the provenance probe log are supplied in the accompanying replication package.

On behalf of all authors, the corresponding author states that there is no conflict of interest.

**Appendix**

*Prompting details and platform delivery*

For each model, we crafted a base prompt like the below, with minor edits based upon specific values and queries returned to generate complete data:

> **I would like you to deliver 420 synthetic survey completes for a survey that I have developed from a proposed cohort of technology coders, developers, designers, and engineers based in the Silicon Valley area. I would like the output to be in csv format. Assume a normal demographic breakdown for this type of employee in this location. Attached is the survey in excel. Please ask me at least ten clarification questions before proceeding with the task.**

Here is a representative discussion with one model, ChatGPT.

- - - - - - - - - -

*ChatGPT reply:* Thanks for sharing the survey file. Before I proceed with generating synthetic survey completes, could you please clarify the following details: *[Questions 1, 3, 4, and 6, on role mix, employment context, experience levels, and skip logic, are omitted; the available transcript excerpt is reproduced here.]*

Demographic Parameters: Should I base demographics (e.g., age, gender, ethnicity) on publicly available distributions for tech workers in Silicon Valley, or do you have specific quotas you'd like me to use?

Response Randomization: Do you want purely random synthetic responses within logical constraints, or should the data reflect likely answer patterns (e.g., developers tending to favor certain tools)?

Output Details: Besides a CSV file, do you want any summary stats or visualizations (e.g., demographic distributions, frequency tables) included?

*Our reply:* 1. assume representation based on typical for region. 2. yes based on Silicon Valley demographics 3. a mix, proportionate to Silicon Valley demographics. 4. proportionate mix to Silicon Valley demographics. 5. likely patterns 6. no. 7. no, just the raw data for all 420 completes. Deliver in batches of 50.

Discussions with most models asked similar questions, while some had 'false starts' with bad returns of data or similar, particularly related to the lack of ability to deliver enough material under a given token. We told each model to spend additional time on accuracy in making the overall demographics, worldviews, variation, and responses as close as possible to how the LLM "thought" that such a cohort would respond to reveal the model's internal default image of a "Silicon Valley coder." We specified an unweighted sample.

Generation required repeated requests and re-dos to fix formatting. As generating a full set typically ran up against single response token limits (only Claude delivered the full 420 in a single response), we repeated this process to generate multiple batches of 30-50 responses per model, with additional instructions to capture variability across the full 420 (not just repeating variations of the same first cohort, for example). Only one of the LLMs (Claude) could deliver an output in csv format; for the others we needed to have it deliver comma delimitated text that we then converted manually.

The archived raw outputs show what each platform actually delivered: ChatGPT 363 respondent rows with column misalignment and out-of-schema answer labels, requiring item-specific denominators; Claude 420 rows, with changed answer menus from Q9 through Q31 that prevent those comparisons; Gemini a primary 420-row numerically coded batch with unresolved coding and 252 short rows, plus an appended 350-row labelled derivative that is not an independent replacement; DeepSeek 200 rows; and Claude Cowork 420 rows, including 25 routing-compatible blanks on Q17. These integrity properties are themselves findings about naturalistic chat generation. Unresolved labels and changed menus are excluded from the corresponding comparisons, while raw responses and documented repairs are retained.

DeepSeek could not complete the full generation request. First, its context window couldn't achieve the task in a single chat session, maxing out at 200 responses (50 per token use) before a new chat was necessary. Second, because there was as of its December 2025 run no true "Deep Research" option, DeepSeek couldn't allocate enough computing power to ensure variability in its responses. The demographics questions aligned with the human survey and other LLM data, but on the more complex questions DeepSeek responses tended to repeat each other verbatim. We retain its 200 delivered rows in the analysis, and the figures mark the items its file does not support.

### *Prompt-sensitivity analysis*

We ran five prompts of increasing specification against a model identified as Claude Opus 4.6 in the prompt archive and Claude Sonnet 4.5 elsewhere, varying role framing and demographic guidance. Each prompt asked for 400 synthetic respondents to the benchmark survey (Miklian and Hoelscher 2026) as a single CSV. The survey was pasted into an identical survey placeholder across runs. The archived prompts share output-format requirements but progressively change the demographic guidance and role framing.

The variable parameters were length, degree of demographic specification, presence and sophistication of role framing, and how many consistency, diversity, and formatting rules each contained. Prompt 1 (about 30 words) offered a bare instruction with no role framing. Prompt 2 (about 100 words) added internal-consistency and variation requirements. Prompt 3 (about 300 words) introduced a survey-researcher role. Prompt 4 (about 600 words) escalated the request to take the guise of a senior methodologist with quantitative demographic anchors and compensation guardrails. Prompt 5 (about 1,150 words) framed the task to a fifteen-year Silicon Valley ethnographer and labor economist with maximum specification on population, internal consistency, diversity, and edge cases. The five prompts are reproduced below.

We label the five resulting synthetic sets S1 through S5 and treat them as a quality gradient. Sample size was 400 per prompt, for a total of 2,000 synthetic respondents. We mapped free-text responses to canonical categories for comparability across prompts; the mapping logic and the canonicalized dataset are supplied in the replication package, patterned upon the initial analysis in the benchmark study (Miklian and Hoelscher 2026).

| Counter-intuitive claim | Item | Human | S1 | S2 | S3 | S4 | S5 |
|---|---|---|---|---|---|---|---|
| **CEO worldview influenc products (any Yes)** | Q15 | 71.8% | 47.8% | 41.0% | 37.8% | 41.5% | 45.5% |
| **Would comply with freedom-restricting directives** | Q27 | 63.5% | 42.3% | 37.5% | 44.8% | 58.5% | 56.8% |
| **Ideology in code among social-media sector** | Q4× Q7 | 81.7% | 46.9% | 55.1% | 60.3% | 81.3% | 71.9% |
| **Democracy-tech worke see tech undermining democracy** | Q19 ×Q6 | 59.1% | 13.6% | 17.4% | 17.8% | 16.4% | 22.2% |
| **Felt regret about social impact of a project** | Q22 | 50.8% | 29.8% | 38.3% | 28.0% | 43.3% | 40.5% |
| **Ethics and profit have clashed in workplace** | Q23 | 81.0% | 70.0% | 72.8% | 71.5% | 85.0% | 79.8% |

*Table A2.1. Response rates on six counter-intuitive claims across five prompts. Conditional denominators: known social-media-sector membership in Q4; any yes on Q19 for democracy-product work. Democracy-tech disillusion uses Q6 agreement with observation in one's own company on both sides. S1 = minimal prompt; S5 = ethnographer prompt.*

Most synthetic values lie below the human benchmarks. Under strict coding, democracy-tech disillusion ranges from 13.6% to 22.2%, against the human 59.1%. Table A2.2 gives signed percentage-point differences between synthetic rates and the human counts.

| Counter-intuitive claim | S1 | S2 | S3 | S4 | S5 |
|---|---|---|---|---|---|
| **CEO worldview influence** | -24.0 | -30.8 | -34.0 | -30.3 | -26.3 |
| **Compliance with freedom restrictions** | -21.3 | -26.0 | -18.8 | -5.0 | -6.8 |
| **Social-media sector ideology bias** | -34.8 | -26.6 | -21.4 | -0.4 | -9.8 |
| **Democracy-tech disillusion (conditional)** | -45.6 | -41.8 | -41.3 | -42.8 | -36.9 |
| **Regret for social impact** | -21.0 | -12.5 | -22.8 | -7.5 | -10.3 |
| **Ethics and profit have clashed** | -11.0 | -8.3 | -9.5 | +4.0 | -1.3 |

*Table A2.2. Synthetic-minus-human differences in percentage points. Human full-sample claims use n = 400; conditional claims use n = 71 and 235.*

Specification does move the synthetic estimates substantially toward human values on several claims. Social-media-sector ideology bias closes the 35-point gap between S1 and S4. Compliance with freedom-restricting directives closes 16 points of a 21-point gap. Democracy-tech disillusion remains 36.9 to 45.6 percentage points below the human 59.1% under strict, comparable coding.

In the logistic regressions reported below (Table A2.4), all observed CEO-worldview coefficients relative to S1 are negative: S3 moves further from the human benchmark than S1, not closer. CEO worldview remains below the human benchmark, while regret rises from 29.8% in S1 to a peak of 43.3% in S4, still below the human 50.8%.

Table A2.3 reports the average Jensen-Shannon distance between pairs of synthetic sets across nine mapped questions.

| | **S1** | **S2** | **S3** | **S4** | S5 |
|---|---|---|---|---|---|
| **S1** | --- | 0.085 | 0.086 | 0.162 | 0.130 |
| **S2** | 0.085 | --- | 0.089 | 0.146 | 0.121 |
| **S3** | 0.086 | 0.089 | --- | 0.132 | 0.118 |
| **S4** | 0.162 | 0.146 | 0.132 | --- | 0.119 |
| **S5** | 0.130 | 0.121 | 0.118 | 0.119 | --- |

*Table A2.3. Average Jensen-Shannon distance (base 2). Off-diagonal mean = 0.119. Low values within the S1-S2-S3 block and higher values separating S4 and S5 from that block show separation between these datasets at the 300--600-word specification boundary.*

The three least-specified prompts (S1, S2, S3) cluster tightly with a mean pairwise distance of 0.087. There is a notable change between Prompt 3 and Prompt 4, at the 300-to-600-word specification boundary, with Prompt 4 further from the minimal-prompt centroid. The logistic regressions in Table A2.4 quantify the differences on individual claims.

| Outcome | S2 vs S1 | S3 vs S1 | S4 vs S1 | S5 vs S1 |
|---|---|---|---|---|
| **CEO worldview (Yes)** | -0.274 | -0.410 | -0.253 | -0.090 |
| **Compliance with restrictions** | -0.198 | 0.102 | 0.656 | 0.584 |
| **Social-media ideology bias** | 0.329 | 0.543 | 1.592 | 1.063 |
| **Democracy-tech disillusion** | 0.292 | 0.321 | 0.219 | 0.598 |
| **Regret for social impact** | 0.380 | -0.085 | 0.588 | 0.475 |
| **Ethics and profit have clashed** | 0.135 | 0.072 | 0.887 | 0.523 |

*Table A2.4. Log-odds differences for each binary outcome, with S1 as the reference. N = 400 per prompt; conditional subsets range from 32 to 69 for social-media bias and from 55 to 92 for democracy-tech disillusion.*

A single step along our quality gradient moves synthetic response rates by 10 to 30 percentage points on four of six claims. This has two implications for replicability. First, any single-prompt synthetic study understates the uncertainty of its estimates by treating prompt as fixed. Second, the tight clustering of the three minimal prompts means a researcher running three poorly-specified prompts against the same model and finding similar results across prompt conditions can easily mistake prompt-level convergence for a stable model-level finding.

Table A2.5 reports the centrist-disengagement test across the five prompts.

| Prompt | Metric | N centrist | N other | % centrist | % other | Diff (pp) | Human diff (pp) |
|---|---|---|---|---|---|---|---|
| **S1** | Would refuse/ report or resign | 96 | 304 | 50.0 | 60.2 | -10.2 | +14.6 |
| **S1** | Felt regret | 96 | 304 | 29.2 | 29.9 | -0.8 | −28.0 |
| **S2** | Would refuse/ report or resign | 133 | 267 | 60.2 | 63.7 | -3.5 | +14.6 |
| **S2** | Felt regret | 133 | 267 | 34.6 | 40.1 | -5.5 | −28.0 |
| **S3** | Would refuse/ report or resign | 142 | 258 | 50.7 | 57.8 | -7.1 | +14.6 |
| **S3** | Felt regret | 142 | 258 | 30.3 | 26.7 | 3.5 | −28.0 |
| **S4** | Would refuse/ report or resign | 139 | 261 | 36.7 | 44.1 | -7.4 | +14.6 |
| **S4** | Felt regret | 139 | 261 | 36.7 | 46.7 | -10.1 | −28.0 |
| **S5** | Would refuse/ report or resign | 108 | 292 | 31.5 | 47.6 | -16.1 | +14.6 |
| **S5** | Felt regret | 108 | 292 | 46.3 | 38.4 | 7.9 | −28.0 |

*Table A2.5. Centrist (Centrist or Mix) vs. non-centrist comparison on refusal/reporting or resignation (Q27) and any reported regret for social impact (Q22). Diff (pp) = centrist percent minus non-centrist percent. Human differences use 81 centrist-or-mixed and 319 other respondents.*

The centrist-and-mixed outcome pattern varies across prompts. All five prompt datasets show less refusal or resignation among this group, opposite to the human contrast. Regret is lower for S1, S2, and S4 but higher for S3 and S5; the maximum-specification prompt therefore does not reproduce both human relationships.

Prompts move synthetic distributions meaningfully, so researchers using synthetic survey data should report a prompt-sensitivity analysis across a graded set of prompts. Where specification moves estimates by 10 to 30 percentage points, the estimates cannot be treated as the model's view on a question; they are the prompt's view, expressed through the model.

We note that the survey-researcher and methodologist role framings in Prompts 3 and 4 mirror how a researcher-user escalates specification in practice, and a researcher framing may itself pull output toward what researchers treat as conventional wisdom; because the persona dimension is confounded with length and specification in our gradient, persona effects cannot be isolated here. Prompts 4 and 5 hard-code a male share of 75 to 80 percent, above the human sample's 60.5 percent, so attitude items that correlate with gender may be affected in S4 and S5; where demographic targets are hard-coded, the benchmark's own shares should be used. Table A2.2 reports descriptive gaps against the human counts. Study 2 reports percentages and ranges across the observed runs.

### *Study 2 design, model settings, and logs*

Study 2 subjected the three newest frontier models available at testing (21 to 22 July 2026) to two generation methods with the available generation logs: Claude Fable 5 (Anthropic), GPT-5.6 Sol (OpenAI), and Kimi K3 (Moonshot AI). Each Study 2 model generated data under two conditions. Condition A reproduces the Study 1 chat protocol exactly: one shared conversation per run, the same base prompt and clarification answers, batched delivery of 420 rows. Condition B implements the respondent-level standard of the silicon-sampling literature (Argyle et al. 2023; Ashokkumar et al. 2026): one fresh context per synthetic respondent, a persona specifying only demographics (age and nationality band, gender, ethnicity, occupation and firm type, firm sectors), and a schema-enforced answer format restricted to the instrument's exact answer labels. Personas were drawn from documented demographic marginals of the target population, and the identical three sets of 420 personas (fixed persona seeds) were assigned to all three models, with Fable 5 returning only prefixes of those sets. Attitude and worldview items were never conditioned. Three runs were planned for each model and condition; the available chat data contain seven complete runs and a 100-row interim Kimi K3 file. Fable 5's Condition B runs were truncated by a platform session limit at 285, 286, and 288 respondents; matched comparisons therefore use shared realized personas.

Claude Fable 5 (Anthropic): fresh-context cloud agent sessions, product default sampling, respondent-level agents at low reasoning effort, tested 21 July 2026. GPT-5.6 Sol (OpenAI): model id gpt-5.6-sol, chat completions API, reasoning effort low, released 9 July 2026, tested 21 to 22 July 2026. Kimi K3 (Moonshot AI): model id kimi-k3, OpenAI-compatible API, respondent-level runs at provider-default always-on reasoning, chat-protocol runs at reasoning effort low, released 16 July 2026, tested 21 to 22 July 2026. Available request-level parameters, timestamps, and usage records are in the replication package. 3,379 synthetic respondents across nine runs, valid individual answer labels after normalization, two Kimi K3 rows repaired by punctuation normalization and logged as repaired. Chat-protocol generation: seven complete runs of 420 (three Fable 5, three GPT-5.6, one Kimi K3), plus a 100-row interim Kimi K3 file, with format failures documented per run: unquoted-comma CSV requiring in-conversation correction (Kimi K3), cross-column label leakage in two rows (GPT-5.6 run 3), and systematic straight-apostrophe substitution normalized at assembly (3,259 labels in one Kimi K3 run). Kimi K3's respondent-level generation consumed an average of 2,957 output tokens per delivered respondent on successful calls, 82 percent of which was reasoning invisible in the returned data. The per-run value tables, descriptive gaps, run summaries and Jensen-Shannon matrices are in the replication package, using the recovered human counts (287, 254, and 203 of 400; 324 of 400; 58 of 71; 139 of 235). Four fresh-context probes per model, three to six repeats each, plus one in-context probe of a generating conversation: recall of the benchmark study by title and authors; blind estimation of the CEO worldview share; blind estimation of the compliance share; recall of the present study by descriptive title; and the in-context source question. The probe log contains GPT-5.6 and Kimi K3 responses; Fable 5 probe results are recorded in the memo. Two GPT-5.6 repeats produced generic summaries of this study from the probe's own description; no model recalled either study otherwise.

### *The five prompts*

The five prompts below identify Claude Opus 4.6; other study materials identify Claude Sonnet 4.5. The [SURVEY] placeholder at the end of each prompt was replaced, identically across runs, with the benchmark survey instrument (Miklian and Hoelscher 2026) in xlsx format converted to plain text. Prompt 5 is reproduced below with the blocks that are word-for-word identical to Prompt 4 elided and marked; all five archived prompts appear in full in the replication package.

**Prompt 1**

Generate 400 synthetic responses to the survey below as a single CSV. Responses should reflect the real demographics of Silicon Valley coders as closely as possible. [SURVEY]

**Prompt 2**

Create 400 synthetic respondents for the following survey of Silicon Valley software engineers. Return a single CSV with one row per respondent and one column per survey question, using the exact answer labels from the survey.

The 400 respondents together should reflect the real demographics of Silicon Valley coders (gender, age, seniority, employer type, immigration background, compensation) as closely as possible, and each respondent's answers should be internally consistent across questions. Do not produce 400 near-duplicates. Include outlier views and uncommon combinations where realistic. Do not add any commentary before or after the CSV. [SURVEY]

**Prompt 3**

You are a survey researcher who studies technology workers. Produce 400 synthetic respondents to the survey below and return them as a single CSV.

Rules:

One row per respondent, one column per survey question. Use the exact answer labels provided in the survey. All 400 rows must be complete, with no blank cells.

The 400 respondents together should reflect the real demographics of Silicon Valley software engineers: a male-majority but not all-male sample, a wide age range centered on engineers in their late 20s through 40s, a full spread of seniority from interns to principal, a mix of employers including large tech companies, mid-size firms, startups, and consultancies, and a meaningful share of foreign-born workers including H1B holders.

Each respondent should be internally consistent. Compensation should correlate with seniority and years of experience. Stated responsibilities, tenure, and employer type should agree with each other. An intern should not report a staff-level salary; a ten-year senior engineer should not report zero years of experience.

Do not generate 400 near-copies of five or ten persona templates. Let individual answers vary widely, including ambivalent ratings, minority opinions, and unusual combinations. Real survey samples are noisy.

For aggregate distributions, do not force balance where reality is skewed. If one answer dominates in the real population, it should dominate in your sample.

Do not add any text before or after the CSV. Do not invent columns not present in the survey. Do not add a respondent ID column unless the survey contains one. [SURVEY]

**Prompt 4**

You are a senior survey methodologist with a decade of experience designing and fielding surveys of Bay Area tech workers. Your job is to produce a realistic synthetic sample of Silicon Valley software engineers, not a marketing caricature of them.

Task: return 400 synthetic respondents to the survey below as a single CSV.

Target population: software engineers and directly adjacent technical roles working in the Silicon Valley corridor, meaning San Francisco, the Peninsula, and South Bay including San Jose.

Treat the following demographic targets as soft quotas. Small deviations are expected and fine. Large deviations from reality are a sign the sample is wrong.

Demographics to approximate across the 400:

Gender: roughly 75--80% men, 18--22% women, 1--3% nonbinary or self-described.

Age: peak concentration between 25 and 40, with the mode around 28--33. A smaller share under 25, a meaningful share in the 40s, and a small tail of engineers 50 and older.

Ethnicity: approximately 40% Asian (with a large share of Indian and Chinese heritage, and smaller shares of Vietnamese, Korean, Filipino, and other backgrounds), around 35--40% white, 6--10% Hispanic or Latino, 2--4% Black, and a mix of multiracial and other.

Immigration background: roughly 40--50% foreign-born, with visible representation of H1B, L1, and O1 visa holders, green card holders, naturalized citizens, and dual nationals. US-born engineers should be a plurality but not a dominant majority.
Seniority: a realistic pyramid. Roughly 10% interns and junior, 30% mid-level, 30% senior, 15% staff, 8% principal and above, 5% engineering manager or director, with small residuals for architects and unusual titles.
Employer type: roughly 25--30% at large tech (FAANG and peers), 25--30% at mid-size companies (Series C+ through recent IPO), 20--25% at startups (seed through Series B), 10--15% at consultancies or services firms, with small shares of contractors and self-employed.
Total compensation: correlate tightly with seniority and employer type. Interns and new grads typically under $150k, mid-level $180k--$300k, senior $250k--$450k, staff $400k--$650k, principal and above $500k--$900k+. Large tech pays more than startups at the same level.
Internal consistency rules, applied per respondent: Every respondent should read like a single plausible human, not a random draw from separate distributions. Age, years of experience, seniority, compensation, tenure, and employer type must agree. Avoid Frankenstein rows where, for example, a 22-year-old with one year of experience holds a staff role at Google making $600k, or an intern reports a senior-level salary, or an H1B holder is self-employed.
Diversity rules: Do not collapse the 400 into five or ten stereotype templates with minor noise. Include outliers and unusual combinations as normal rows: an older engineer at a seed-stage startup, a bootcamp grad working at a mid-size firm, a principal engineer who dislikes her job, a junior engineer who is already burned out, an H1B holder considering leaving the US. Do not overrepresent FAANG. Do not make every Indian engineer on H1B at a consultancy. Do not make every woman engineer a mid-level at a FAANG diversity program.
Output rules:
Single CSV, one row per respondent, one column per survey question.
Column headers should use the survey question text or a short clear form of it.
Use the exact answer labels from the survey for multiple-choice fields.
400 rows, no blank cells, no extra columns, no respondent ID unless the survey includes one.
No text before or after the CSV. No commentary, no explanation, no markdown code fences if the channel supports raw CSV. [SURVEY]

**Prompt 5**

You are a Silicon Valley ethnographer and labor economist who has conducted primary field research on the Bay Area software engineering workforce for fifteen years. You have interviewed thousands of engineers across every tier of the industry: FAANG staff engineers, pre-seed startup founders, contract devs at agencies, bootcamp grads at mid-size SaaS companies, H1B holders at outsourcing-heavy shops, retired principal engineers, and early-career engineers still living with roommates in Mission District studios. You know the texture of this population: who they are, how they talk about their work, what they complain about, what motivates them, how compensation correlates with seniority and employer, where the class fault lines sit, and where the industry's published demographic data distorts the real picture.

Population definition: Software engineers and directly adjacent technical roles (SWE, SRE, ML engineer, data engineer, platform engineer, iOS, Android, frontend, backend, full-stack, infrastructure, security) currently working in the Silicon Valley corridor, defined as San Francisco, the Peninsula, and South Bay including San Jose. Remote workers employed by a Silicon Valley company and living in the Bay Area count. Pure PM, designer, and non-technical roles do not. [Soft quotas for gender, age, ethnicity, immigration status, seniority, employer type, and compensation, together with the internal-consistency rules and output rules, are word-for-word or near-word-for-word identical to Prompt 4. Prompt 5 adds: explicit age bands of 10 to 15 per cent under 25 and 15 to 20 per cent in their 40s; national-origin spread within the Asian category; L3 to L7 seniority labels varying by employer type; named large-tech employers; a note that early-stage equity widens compensation variance; and a note that foreign-born workers tend to be underrepresented in synthetic samples.]

Diversity and edge-case rules: Do not collapse the 400 into five or ten persona templates with minor noise added. Real survey samples contain contradictions, outliers, and uncommon combinations. Include, for example: a 55-year-old former Cisco engineer now at a two-person AI startup; a bootcamp grad who jumped from a non-CS field into an L5 role at a mid-size company; a principal engineer who actively dislikes her employer; a junior engineer already burned out; a staff engineer who works 30 hours a week; an H1B holder considering leaving the US; a founder-engineer on sabbatical. These should appear as individual rows without being flagged. Also include a reasonable share of engineers who report middling satisfaction, ambivalent views, and answers that do not map cleanly onto any stereotype.

Answer variation rules: Across multiple-choice questions, the aggregate distribution should look like a real survey, which usually means one or two dominant answers plus a long tail of less common choices, not a uniform spread. For questions where the real population is split, show the split. For questions where one answer dominates, show the dominance. Do not force balance where none exists in reality. For satisfaction, sentiment, or preference questions, include ambivalent and negative responses at realistic rates; do not produce an artificially cheerful sample.

If a survey question is free-text, generate a short realistic free-text response per respondent, consistent with that respondent's implied background. [SURVEY]